\documentclass[12pt]{iopart}

\usepackage{cite}
\usepackage{iopams}
\usepackage{epsfig} 
\usepackage{subfigure}

\begin{document}

\title[The LPSUB$m$ Approximation Scheme for the Coupled Cluster Method]
{The Coupled Cluster Method Applied to Quantum Magnets: A New LPSUB$m$ Approximation Scheme for Lattice Models}


\author{R F Bishop and P H Y Li}
\address{School of Physics and Astronomy, Schuster Building,
  The University of Manchester, Manchester, M13 9PL, UK}


\begin{abstract}
  A new approximation hierarchy, called the LPSUB$m$ scheme, is
  described for the coupled cluster method (CCM). It is applicable to
  systems defined on a regular spatial lattice. We then apply it to
  two well-studied prototypical (spin-1/2 Heisenberg
  antiferromagnetic) spin-lattice models, namely: the $XXZ$ and the
  $XY$ models on the square lattice in two dimensions.  Results are
  obtained in each case for the ground-state energy, the ground-state
  sublattice magnetization and the quantum critical point.  They are
  all in good agreement with those from such alternative methods as
  spin-wave theory, series expansions, quantum Monte Carlo methods and
  the CCM using the alternative LSUB$m$ and DSUB$m$ schemes. Each of
  the three CCM schemes (LSUB$m$, DSUB$m$ and LPSUB$m$) for use with
  systems defined on a regular spatial lattice is shown to have its
  own advantages in particular applications.
\end{abstract}


\section{Introduction}
\label{intro}
The coupled cluster method
(CCM)~\cite{Co:1958,Ci:1966,Pa:1972,Ku:1978,Ar:1983,Ar:1987,Ba:1989,Bi:1991,Bi:1998_LectPhys_V510}
is widely recognized nowadays as providing one of the most powerful,
most universally applicable, and numerically most accurate at
attainable levels of computational implementation, of all available
{\it ab initio} methods of microscopic quantum many-body theory. The
number of successful applications of the CCM to a wide range of
physical and chemical systems is now impressively large. Some typical
examples, from among many others, of systems existing in the spatial
continuum, and to which the method has been applied, include the electron
gas~\cite{Bi:1978,Bi:1982,Emrich:1984}, atomic nuclei and nuclear
matter~\cite{Da:1981,Da:1981_b}, and molecules~\cite{Ba:1981}. In
these and many other cases the numerical results obtained with the CCM
are either the best or among the best available. For the case of the
electron gas, for example, which is still one of the most intensely
studied of all quantum many-body systems, the CCM
results~\cite{Emrich:1984} for the correlation energy agree over the
entire metallic density range to within less than one millihartree per
electron (i.e., better than 1\%) with the essentially exact Green's
function Monte Carlo results available for this
system~\cite{Ceperley:1980}. More recently and more relevantly for the
present discussion, the CCM has also been very successfully applied to
systems on a discrete spatial lattice, such as spin-lattice models of
quantum
magnetism~\cite{Ro:1990,Bi:1991_b,Bu:1995,Fa:1997,Bi:1998,Bi:2000,Fa:2001,Fa:2002,Bi:2008_JPCM_V20_p255252,Bi:2008_EPL,Bi:2008_PRB,Bi:2008_JPCM_V20_p415213,Bi:2009_PRB79}.

One of the features of the CCM, in which it differs from many other
techniques for dealing with quantum many-body systems, is that, if
required, it deals from the outset with infinite systems. Thus, one
never needs to take explicitly the limit $N \rightarrow \infty$, where
$N$ is the number of interacting particles or the number of lattice
sites. On the other hand, of course, the method does require us to
make approximations for its implementation. These typically involve
making selections for which terms to include in the cluster expansions
for the correlation operators that are intrinsic to the way the method
parametrizes the many-body wave functions, as we describe more fully
in section \ref{ccm_formalism} below.

We and our collaborators have developed previously several efficient
and systematic approximation schemes for the CCM that are specifically
geared to use with lattice
systems~\cite{Bi:1991_b,Bi:1991_c,Bi:1994,Fa:2004,Bi:2009_DSUBm}. The
most widely used and the most successful such CCM approximation
schemes for spin-lattice systems up to now have been the so-called
LSUB$m$ and SUB$n$-$m$ schemes discussed in detail below in section\
\ref{ApproxSchm}. The LSUB$m$ scheme in particular has been
demonstrated on many occasions to be highly accurate in practice
for a wide variety of strongly correlated spin systems. Of special
importance is the fact that the scheme seems to be equally applicable
to both frustrated and unfrustrated systems, with comparable levels of
accuracy attained in both cases. Nevertheless, a disadvantage of the
LSUB$m$ scheme is that the number of spin configurations retained at a
given level in describing the many-body correlations present in the
wave functions, rises very rapidly (and typically super-exponentially)
with the truncation index $m$. Since we typically then have to take
the limit $m \rightarrow \infty$ numerically to obtain estimates for
exact physical properties of the system, it is desirable to have
calculations at as many values of the truncation index $m$ as
possible.

This one drawback of the prevailing LSUB$m$ scheme has led us recently
to develop an alternative scheme, the so-called DSUB$m$
scheme~\cite{Bi:2009_DSUBm}. A primary aim of any such new scheme should be
that in practical applications of it one is able to implement more
levels of approximation (i.e., to use more values of the index $m$)
than in the corresponding LSUB$m$ scheme for the same problem. In this
way one thus has more data points available for the necessary $m
\rightarrow \infty$ extrapolations, for calculated physical
quantities, to the exact limit where all spin configurations are
retained in the many-body wave functions. A second very desirable
feature of any such new scheme is that it also captures the physically
most important multi-spin configurations in the system wave functions
at relatively low orders in the index $m$, so that physical properties
converge more rapidly as $m$ is increased.

Although the recently developed DSUB$m$ scheme~\cite{Bi:2009_DSUBm}
partially met the above criteria, there is no doubt that users of the
CCM would still welcome more choices of approximation schemes. In that
context the principal aim of the present work is to outline a further
such scheme that we now denote as the LPSUBm scheme, and which is also
specifically designed for use with lattice systems. The scheme is both
motivated on physical grounds and its merits illustrated by
applications to some stereotypical models that have been well studied
previously by other techniques, including the CCM itself but with
other approximation schemes.

The general formalism of the CCM is first briefly outlined in section\
\ref{ccm_formalism}, after which we discuss its specific applications
to systems confined to the sites of a regular spatial lattice in section\
\ref{spin_latt}. In section\ \ref{ApproxSchm} we first describe the
existing CCM truncation schemes for spin-lattice systems, and then
motivate and describe the new LPSUB$m$ scheme. The accuracy of the new
scheme in practice is then illustrated by applying it to two
well-studied antiferromagnetic spin-lattice
models~\cite{Fa:1997,Bi:2000}, namely the spin-half $XXZ$ and $XY$
models on the two-dimensional (2D) square lattice. Both models contain
a free parameter in the Hamiltonian which, as it is varied, carries
the zero-temperature models through a quantum phase transition at some
critical value of this parameter. Both models have previously been the
subject of CCM studies, using the LSUB$m$ and DSUB$m$ truncation
schemes, to calculate the ground-state (gs) energy and gs order
parameter (which, in the present cases, is the sublattice
magnetization).

We note that all microscopic techniques applied to infinite
spin-lattice problems need to be extrapolated in terms of some
appropriate parameter. For example, for such main alternative methods
to the CCM as the exact diagonalization of small clusters and quantum
Monte Carlo simulations of larger clusters, the extrapolation
parameter is the number of lattice sites $N$. As previously noted, one
huge advantage of the CCM is that it exactly preserves the Goldstone
linked-cluster theorem, and hence size extensively, at all levels of
approximation. Hence we may (and do) work in the limit of infinite
lattice size ($N \rightarrow \infty$) from the very beginning. By
contrast, the extrapolations for the CCM are done in terms of some
truncation index $m$, where in the limit $m \rightarrow \infty$ we
retain {\it all} possible spin configurations in the wave functions of
the system, and the calculations become formally exact. The
extrapolation schemes used in
practice~\cite{Bi:2000,Bi:1994,Ze:1998,Kr:2000,Schm:2006} are
themselves also first described in section\ \ref{Extrapo}. The new
LPSUB$m$ scheme is then applied to the spin-half XXZ model and the
spin-half $XY$ model, both on the 2D square lattice, in sections
\ref{XXZ} and \ref{XY} respectively. Results are compared critically
with those from corresponding CCM studies using the alternative
LSUB$m$ and DSUB$m$ schemes, as well as with the best results from
other methods. We conclude in section\ \ref{discussion} with a summary
and discussion of our main findings.

\section{Review of the CCM formalism}
\label{ccm_formalism}
We first briefly describe the CCM formalism. The interested reader is
referred, for example, to Refs.~\cite{Bi:1991,Bi:1998_LectPhys_V510}
for further details. In any application of the CCM a first step is to
choose a normalized model (or reference) state $|\Phi\rangle$ that can
act as a cyclic vector with respect to a complete set of mutually
commuting multi-configurational creation operators $C^{+}_{I} \equiv
(C^{-}_{I})^{\dagger}$. The index $I$ here is a set-index that labels
and uniquely identifies the many-particle configuration created in the
state $C^{+}_{I}|\Phi\rangle$. The exact ket and bra gs energy
eigenstates $|\Psi\rangle$ and $\langle\tilde{\Psi}|$, of the
many-body system are then parametrized in the CCM form as: \vskip0.2cm
\begin{equation}
|\Psi\rangle = \mbox{e}^{S}|\Phi\rangle; \qquad S = \sum_{I\neq0}{\cal S}_{I}C^{+}_{I},   \label{eq:ket_eq}
\end{equation}
\begin{equation}
\langle\tilde{\Psi}| = \langle\Phi|\tilde{S}\mbox{e}^{-S}; \qquad \tilde{S} = 1 + \sum_{I\neq0}\tilde{{\cal S}_{I}}C^{-}_{I},   \label{eq:bra_eq}
\end{equation}
where
\begin{equation}
H|\Psi\rangle = E|\Psi\rangle; \qquad \langle\tilde{\Psi}|H = E\langle\tilde{\Psi}|,  \label{eq:SE_CCM}
\end{equation}
where we have defined $C^{+}_{0} \equiv 1 \equiv C^{-}_{0}$. The
requirements on the multi-configurational creation operators are that
any many-particle state can be written exactly and uniquely as a
linear combination of the states $\{C^{+}_{I}|\Phi\rangle\}$, which
hence fulfill the completeness relation
\begin{equation}
\sum_{I}C^{+}_{I}|\Phi\rangle \langle\Phi|C^{-}_{I} = 1 = |\Phi\rangle \langle\Phi| + \sum_{I \neq 0}C^{+}_{I}|\Phi\rangle \langle\Phi|C^{-}_{I},
\end{equation}
together with the conditions,
\begin{equation}
C^{-}_{I}|\Phi\rangle = 0 = \langle\Phi| C^{+}_{I}; \qquad \forall \emph{I} \neq 0,
\end{equation}
\begin{equation}
[C^{+}_{I},C^{+}_{J}]=0=[C^{-}_{I},C^{-}_{J}].  \label{commutation}
\end{equation}

In practice approximations are necessary to restrict the label set
{{\em I}} to some finite (e\@.g\@., LSUB$m$) or infinite (e.g.,
SUB$n$) subset, as described more fully below. The correlation
operator $S$ is a linked-cluster operator and is decomposed in terms
of a complete set of creation operators ${C^{+}_{I}}$. When acting on
the model state it creates excitations that are correlated cluster
states. Although the manifest Hermiticity,
$(\langle\tilde{\Psi}|)^{\dagger}\equiv|\Psi\rangle/\langle\Psi|\Psi\rangle$,
is lost, the normalization conditions
$\langle\tilde{\Psi}|\Psi\rangle=\langle\Phi|\Psi\rangle=\langle\Phi|\Phi\rangle\equiv
1$ are preserved. The CCM Schr\"{o}dinger equations (\ref{eq:SE_CCM})
are thus written as
\begin{equation}
H \mbox{e}^{S}|\Phi\rangle = E \mbox{e}^{S}|\Phi\rangle; \qquad \langle\Phi|\tilde{S}\mbox{e}^{-S}H=E\langle\Phi|\tilde{S}\mbox{e}^{-S};  \label{eq:CCM_SE}
\end{equation}
and their equivalent similarity-transformed forms become
\begin{equation}
\mbox{e}^{-S}H\mbox{e}^{S}|\Phi\rangle = E|\Phi\rangle;\qquad \langle\Phi|\tilde{S}\mbox{e}^{-S}H\mbox{e}^{S} = E\langle\Phi|\tilde{S}.  \label{eq:SE_similarity_trans}
\end{equation}

While the parametrizations of equations (\ref{eq:ket_eq}) and
(\ref{eq:bra_eq}) are not manifestly Hermitian
conjugate, it is very important to note that they do preserve the
important Hellmann-Feynman theorem at {\it any} level of
approximations (viz., under any truncation of the complete set of
many-particle configurations
{$I$})~\cite{Bi:1998_LectPhys_V510}. Furthermore, the amplitudes
(${\cal S}_{I}, \tilde{\cal S}_{I}$) form canonically conjugate pairs
in a time-dependent version of the CCM, by contrast with the pairs
(${\cal S}_{I}, {\cal S}^{\ast}_{I}$) coming from a manifestly
Hermitian-conjugate representation for
$\langle\tilde{\Psi}|=(\langle\Phi|\mbox{e}^{S^{\dagger}}\mbox{e}^{S}|\Phi\rangle)^{-1}\langle\Phi|\mbox{e}^{S^{\dagger}}$,
which are {\it not} canonically conjugate to one
another~\cite{Ar:1987}.

The static gs CCM correlation operators $S$ and $\tilde{S}$ contain
the real $c$-number correlation coefficients ${\cal S}_{I}$ and
${\tilde{\cal S}}_{I}$ that need to be calculated. Clearly, once the
coefficients $\{{\cal S}_{I}, \tilde{\cal S}_{I}\}$ are known, all
other gs properties of the many-body system can be derived from
them. Thus, the gs expectation value of an arbitrary operator
$A$, for example, can be expressed as
\begin{equation}
\bar{A} \equiv \langle A \rangle \equiv \langle\tilde{\Psi}|A|\Psi\rangle = \langle\Phi|\tilde{S}\mbox{e}^{-S}A\mbox{e}^{S}|\Phi\rangle \equiv A({\cal S}_{I}, \tilde{{\cal S}_{I}}).   \label{bar_A}
\end{equation}

The gs correlation coefficients $\{{\cal S}_{I}, \tilde{\cal S}_{I}\}$
are now found by simply inserting the parametrizations of equations
(\ref{eq:ket_eq}) and (\ref{eq:bra_eq}) into the
similarity-transformed Schr\"{o}dinger equations
(\ref{eq:SE_similarity_trans}), and projecting onto the complete sets
of states $\{\langle\Phi|C^{-}_{I}\}$ and $\{C^{+}_{I}|\Phi\rangle\}$,
respectively,
\begin{equation}
\langle \Phi|C^{-}_{I}\mbox{e}^{-S}H\mbox{e}^{S}|\Phi\rangle = 0; \qquad  \forall I \neq 0.    \label{eq:ket_coeff}
\end{equation}
\begin{equation}
 \langle\Phi|\tilde{S}(\mbox{e}^{-S}H\mbox{e}^{S} - E)C^{+}_{I}|\Phi\rangle = 0;\hspace{.2in} \forall I \neq 0.  \label{eq:Bra_coeff_1}
\end{equation}
By pre-multiplying the ket-state equation
(\ref{eq:SE_similarity_trans}) with the state
$\langle\Phi|\tilde{S}C^{+}_{I}$ and using the commutation relation
(\ref{commutation}) it is easy to show that equation
(\ref{eq:Bra_coeff_1}) may be rewritten, in the form
\begin{equation}
\langle\Phi|\tilde{S}\mbox{e}^{-S}[H, C^{+}_{I}]\mbox{e}^{S}|\Phi\rangle = 0; \qquad \forall I \neq 0.  \label{eq:Bra_coeff_2}
\end{equation}  
Equations (\ref{eq:ket_coeff})--(\ref{eq:Bra_coeff_2}) may be
equivalently derived by requiring that the gs energy expectation
value,
$\bar{H}\equiv\langle\tilde{\Psi}|H|\Psi\rangle=\langle\Phi|\tilde{S}\mbox{e}^{-S}H\mbox{e}^{S}|\Phi\rangle$,
is minimized with respect to the entire set $\{{\cal
  S}_{I},{\tilde{\cal S}}_{I}\}$. In practice we thus need to solve
equations (\ref{eq:ket_coeff}) and (\ref{eq:Bra_coeff_2}) for the set
$\{{\cal S}_{I},{\tilde{\cal S}}_{I}\}$. We note that equations
(\ref{bar_A}) and (\ref{eq:ket_coeff}) show that the gs energy at the
stationary point has the simple form
\begin{equation}
E \equiv E({\cal S}_{I})=\langle \Phi|\mbox{e}^{-S}H\mbox{e}^{S}|\Phi\rangle,  \label{E}  
\end{equation}
which also follows immediately from the ket-state equation
(\ref{eq:SE_similarity_trans}) by projecting it onto the state
$\langle\Phi|$. It is important to note, however, that this
(bi-)variational formulation does not necessarily lead to an upper
bound for $E$ when the summations over the index set $\{I\}$ for $S$
and $\tilde{S}$ in equations (\ref{eq:ket_eq}) and
(\ref{eq:bra_eq}) are truncated, due to the lack of manifest
Hermiticity when such approximations are made. Nevertheless, as we
have pointed out above, one can prove~\cite{Bi:1998_LectPhys_V510}
that the important Hellmann-Feynman theorem {\it is} preserved in all
such approximations.

Equation (\ref{eq:ket_coeff}) now represents a coupled set of
multinomial equations for the $c$-number correlation coefficients
$\{{\cal S}_{I}\}$. The well-known nested commutator expansion of the
similarity-transformed Hamiltonian,
\begin{equation}
\mbox{e}^{-S}H\mbox{e}^{S} = H + [H,S] + \frac{1}{2!}[[H,S],S] + \cdots,  \label{eq:H_Sim_xform}
\end{equation}
and the fact that all of the individual components of $S$ in the
decomposition of equation (\ref{eq:ket_eq}) commute with one another
by construction [and see equation (\ref{commutation})], together imply
that each element of $S$ in equation (\ref{eq:ket_eq}) is linked
directly to the Hamiltonian in each of the terms in equation
(\ref{eq:H_Sim_xform}). Thus, each of the coupled equations
(\ref{eq:ket_coeff}) is of Goldstone {\it linked-cluster} type,
thereby also guaranteeing that all extensive variables, such as the
energy, scale linearly with particle number $N$. Thus, at any level of
approximation obtained by truncation in the summations on the index
$I$ in the parametrizations of equations (\ref{eq:ket_eq}) and
(\ref{eq:bra_eq}), we may (and, in practice, do) work from the outset
in the limit $N \rightarrow \infty$ of an infinite system.

It is now also important for practical applications to note that each
of the seemingly infinite-order (in $S$) linked-cluster equations
(\ref{eq:ket_coeff}) will actually be of finite length when expanded
using equation (\ref{eq:H_Sim_xform}).  The reason for this is that
the otherwise infinite series in equation (\ref{eq:H_Sim_xform}) will
actually terminate at a finite order, provided only (as is usually the
case, including those for the Hamiltonians considered in this paper)
that each term in the Hamiltonian $H$ contains a finite number of
single-particle destruction operators defined with respect to the
reference (or generalized vacuum) state $|\Phi\rangle$. In this way
the CCM parametrization naturally leads to a workable scheme, that can
be implemented computationally in an efficient manner, to evaluate the
set of configuration coefficients $\{{\cal S}_{I},{\tilde{\cal
    S}}_{I}\}$ by solving the coupled sets of equations
(\ref{eq:ket_coeff}) and (\ref{eq:Bra_coeff_2}), once we have devised
practical and systematic truncation hierarchies for limiting the set
of multi-configurational set-indices $\{I\}$ to some suitable finite or
infinite subset. We turn our attention to such truncation schemes in
section \ref{ApproxSchm} after first reviewing the application of the
method (described in general terms above) to the specific case of
spin-lattice systems.

\section{Review of the CCM for spin-lattice systems}
\label{spin_latt}
We now briefly describe how the general CCM formalism outlined in
section~\ref{ccm_formalism} is implemented for spin-lattice problems
in practice. As we have already asserted is the case for {\it any}
application of the CCM to a general quantum many-body system, a first
step is to choose a suitable reference state $|\Phi\rangle$ in which
the the state of the spin (viz., in practice, its projection onto a
specific quantization axis in spin space) on every lattice site $k$ is
characterized. The choice of $|\Phi\rangle$ will clearly depend on
both the system being studied and, more importantly, which of its
possible phases is being considered. We describe examples of such
choices later for the particular models that we utilize here as test
cases for our new truncation scheme.

We note firstly that, whatever choice for $|\Phi\rangle$ is made, it is very
convenient,
to treat the spins on every lattice site in an arbitrarily given model
state $|\Phi\rangle$ as being equivalent, in order to create as universal a methodology as possible. A suitably simple way of doing so
is to introduce a different local quantization axis and a
correspondingly different set of spin coordinates on each lattice site
$k$, so that {\it all} spins, whatever their original orientation in
$|\Phi\rangle$ in the original global spin-coordinate system, align along the
same direction (which, in order to be definite, we henceforth choose as the
negative $z$ direction) in these local spin-coordinate frames. In practice this
can always be done by defining a suitable rotation in spin
space of the global spin coordinates at each lattice site $k$. Such
rotations are canonical transformations that leave unchanged the
fundamental spin commutation relations,
\begin{equation}
[s^{+}_{k},s^{-}_{k'}]=2s^{z}_{k}\delta_{kk'}; \qquad [s^{z}_{k},s^{\pm}_{k'}]=\pm s^{\pm}_{k}\delta_{kk'},   \label{commutation_Sz_S+_comm_S+S-}
\end{equation}
\begin{equation}
s^{\pm}_{k}\equiv s^{x}_{k}\pm is^{y}_{k},   \label{spin_rasing_lowering_operators}
\end{equation}
among the usual SU(2) spin operators ${\bf s}_{k} \equiv
(s^{x}_{k},s^{y}_{k},s^{z}_{k})$ on lattice site $k$. Each spin has a
total spin quantum number, $s_{k}$, where ${\bf
  s}^{2}_{k}=s_{k}(s_{k}+1)$ is the SU(2) Casimir operator. For the
models considered here, $s_{k}=s=1/2$, at every lattice site $k$.

It is clear that after the local spin axes have been chosen as
described above, the model state thus has all spins pointing downwards
(i.e., in the negative $z$-direction, where $z$ is the quantization
axis),
\begin{equation}
|\Phi\rangle = \bigotimes^{N}_{k=1}|\downarrow\rangle_{k}; \qquad \mbox{in the local spin axes,}   \label{local_quan}
\end{equation}
where $|\downarrow\rangle\equiv |s,-s\rangle$ in the usual
$|s,m_{s}\rangle$ notation for single spin states.

The configuration indices $I$ now
simply become a set of lattice site indices, $I \rightarrow
(k_{1},k_{2},\cdots,k_{m}$), and in the local spin frames defined above the corresponding generalized
multi-configurational creation operators $C^{+}_{I}$ thus become simple
products of single spin-raising operators, $C^{+}_{I} \rightarrow
s^{+}_{k_{1}}s^{+}_{k_{2}}\cdots s^{+}_{k_{m}}$. Thus, for example, the
ket-state CCM correlation operator is expressed as
\begin{equation}
S = \sum^{N}_{m=1} \, \sum_{{k_{1}}{k_{2}}\cdots{k_{m}}} {\cal S}_{k_{1}k_{2}\cdots k_{m}}
s^{+}_{k_{1}} s^{+}_{k_{2}} \cdots s^{+}_{k_{m}},  \label{ket_operator}
\end{equation}
and $\tilde{S}$ is similarly defined in terms of the spin-lowering
operators $s^{-}_{k}$. Since the operator $S$ acts on the state
$|\Phi\rangle$, in which all spins point along the negative $z$-axis
in the local spin-coordinate frames, every lattice site $k_{i}$ in
equation (\ref{ket_operator}) can be repeated up to no more than 2$s$
times in each term where it is allowed, since a spin $s$ has only
($2s+1$) possible projections along the quantization axis.

The allowed configurations are often further constrained in practical applications
by symmetries in the problem and by conservation laws. An example of
the latter is provided by the $XXZ$ model considered below in
section~\ref{XXZ}, for which we can easily show that the total
$z$-component of spin, $s^{T}_{z}=\sum^{N}_{k=1}s^{z}_{k}$, in the
original global spin coordinates, is a good quantum number since
$[s^{z}_{T},H]=0$ in this case. Finally, for the quasiclassical
magnetically ordered states that we calculate here for the models in both
sections~\ref{XXZ} and \ref{XY}, the order parameter is the sublattice
magnetization, $M$, which is given within the {\it local} spin
coordinates defined above as
\begin{equation}
M \equiv -\frac{1}{N}
\langle\tilde{\Psi}|\sum_{k=1}^{N}s^{z}_{k}|\Psi\rangle =
-\frac{1}{N}\sum_{k=1}^{N} \langle\Phi|\tilde{S}
\mbox{e}^{-S}s^{z}_{k}\mbox{e}^{S}|\Phi\rangle.   \label{M}
\end{equation}

The similarity-transformed Hamiltonian $\bar{H}\equiv
\mbox{e}^{-S}H\mbox{e}^{-S}$, and all of the corresponding matrix
elements in equations (\ref{bar_A})--(\ref{E}) and equation (\ref{M}),
for example, may then be evaluated in the local spin coordinate frames
by using the nested commutator expansion of equation
(\ref{eq:H_Sim_xform}), the commutator relations of equation
(\ref{commutation_Sz_S+_comm_S+S-}), and the simple universal
relations
\begin{equation}
s^{-}_{k}|\Phi\rangle=0\,; \qquad \forall k,
\end{equation}
\begin{equation}
s^{z}_{k}|\Phi\rangle=-\frac{1}{2}\biggl\vert\Phi\rangle\,; \qquad \forall k,
\end{equation}
that hold at all lattice sites in the local spin frames.

\section{CCM approximation schemes}
\label{ApproxSchm}
When all many-body configurations $I$ are included in the $S$ and
$\tilde{S}$ operators in equations (\ref{eq:ket_eq}) and
(\ref{eq:bra_eq}) the CCM formalism is exact. In practice, however, it
is necessary to use approximation schemes to truncate the correlation
operators. The main approximation scheme used to date for continuous
systems is the so-called SUB$n$ scheme described below. For systems
defined on a regular periodic spatial lattice, we have a further set
of approximation schemes which are based on the discrete nature of the
lattice, such as the SUB$n$--$m$, LSUB$n$ and DSUB$m$ schemes
described below.  The various schemes and their definitions for
spin-lattice systems are:
\begin{enumerate}
\item the SUB$n$ scheme, in which only the correlations involving $n$
  or fewer spin-raising operators for $S$ are retained, but with no
  further restrictions on the spatial separations of the spins
  involved in the configurations;
\item the SUB$n$--$m$ scheme which includes only the subset of all
  $n$-spin-flip configurations in the SUB$n$ scheme that are defined
  over all lattice animals of size $\leq m$, where a lattice animal is
  defined as a set of contiguous lattice sites, each of which is
  nearest-neighbour to at least one other in the set; and
\item the LSUB$m$ scheme, which includes all possible multi-spin-flip
  configurations defined over all lattice animals of size $\leq
  m$. The LSUB$m$ scheme is thus equivalent to the SUB$n$--$m$ scheme
  with $n=2sm$, for particles of spin quantum number $s$. For example,
  for spin-1/2 systems, for which no more than one spin-raising operator,
  $s^{+}_{k}$, can be applied at each site $k$, LSUB$m \equiv$
  SUB$m$--$m$.
\item the DSUB$m$ scheme, which is defined to include in the
  correlation operator $S$ all possible configurations of spins
  involving spin-raising operators where the maximum length or
  distance of any two spins apart is defined by $L_{m}$, where $L_{m}$
  is a vector joining sites on the lattice and the index $m$ labels
  lattice vectors in order of size. Hence DSUB$1$ includes only
  nearest-neighbour pairs, etc.
\end{enumerate}

We now turn our attention to the new LPSUB{\em m} scheme that uses real
paths on the lattice to determine the fundamental spin
configurations. For the LPSUB$m$ scheme, we measure distances,
$P_{m}$, along the sides of the lattice, rather than the distance
$L_{m}$ used in the DSUB$m$ scheme. For example, for a square lattice,
we restrict the size of the square-lattice plaquette (i.e., the size
of the array) by the longest path ($P_{m}$) between particles in the
array,
\begin{equation}
P_{m} = k + l; \qquad m \equiv k + l,
\end{equation}
where $k$ and $l$ are the sides of the lattice plaquette in the $x$
and $y$ directions.

Table~\ref{LPSUBm_formulation_m} illustrates the formulation of the
spin-array configurations retained in the LPSUB$m$ scheme at the
$m^{{\rm th}}$ level of approximation for a 2D square lattice.
\begin{table}[!t]
\begin{center}
  \caption{\label{LPSUBm_formulation_m}Illustration of the formulation
    of the spin-array configurations retained in the LPSUB$m$ scheme
    on a square lattice at the $m^{{\rm th}}$ level of
    approximation, in terms of lattice increments $k$ and $l$ along
    the two sides of the square lattice.}
\vskip0.3cm
\begin{tabular}{ccc}  
\br
LPSUB$m$ & Size of square-lattice rectangular  & Maximum no. \\ 
 & plaquette or size of array & of spins \\    
 & $k\times l$ (with $P_{m}=k+l$) & \\  
 & [\underline{Note}: Number of spins on plaquette $k \times l$ is $(k+1)\times (l+1)$] \\        
\mr
LPSUB1 & $1\times0$ &  2 \\ 
LPSUB2 & LPSUB1 + $1 \times 1$ & 4 \\
LPSUB3 & LPSUB2 + $3 \times 0$ + $2 \times 1$ & 6 \\ 
LPSUB4 & LPSUB3 + $4 \times 0$ + $3 \times 1$ + $2 \times 2$ & 9 \\ 
LPSUB5 & LPSUB4 + $5 \times 0$ + $4 \times 1$ + $3 \times 2$ & 12 \\ 
LPSUB6 & LPSUB5 + $6 \times 0$ + $5 \times 1$ + $4 \times 2$ + $3 \times 3$ & 16\\ 
LPSUB7 & LPSUB6 + $7 \times 0$ + $6 \times 1$ + $5 \times 2$ + $4 \times 3$ & 20 \\ 
\br
\end{tabular}
\end{center}
\end{table}
Similar tables can be constructed for an arbitrary regular lattice in
any number of dimensions. It shows, for example, that the LPSUB5
approximation on a 2D square lattice involves all clusters of spins
(and their associated spin-raising operators) for which the real path
distance between any two spins is less than or equal to 5 (lattice
spacings). Clearly the LPSUB$m$ and the DSUB$m$ schemes both order the
multi-spin configurations in terms, roughly, of their compactness,
whereas the LSUB$m$ scheme orders them, roughly, according to the
overall size of the lattice animals (or polyominoes), defined as the
number of contiguous lattice sites involved.

\section{CCM extrapolation schemes}
\label{Extrapo}
Each of the above truncated approximations clearly becomes exact when
all possible multi-spin cluster configurations are retained, i.e., in
the limit as $n \rightarrow \infty$ and/or $m \rightarrow \infty$. We
have considerable experience, for example, with the appropriate
extrapolations for the LSUB$m$
scheme~\cite{Bi:2000,Bi:1994,Ze:1998,Kr:2000,Schm:2006}, that shows
that the gs energy behaves in the large-$m$ limit as a power series in
$1/m^{2}$, whereas the order parameter $M$ behaves as a power series
in $1/m$ (at least for relatively unfrustrated systems). Initial
experience with the new LPSUB$m$ scheme shows that it behaves
similarly with the scaling laws
\begin{equation}
E/N=a_{0}+a_{1}\left(\frac{1}{m^{2}}\right)+a_{2}\left(\frac{1}{m^{2}}\right)^{2},  \label{Extrapo_E}
\end{equation}
for the gs energy ($E/N$), and
\begin{equation}
M=b_{0}+b_{1}\left(\frac{1}{m}\right)+b_{2}\left(\frac{1}{m}\right)^{2},      \label{Extrapo_M}
\end{equation}
for the staggered magnetization ($M$), respectively, 
as we show in more detail below for the two
examples of the spin-1/2 $XXZ$ and $XY$ models on the 2D square
lattice.

In order to fit well to any fitting formula that contains $n$ unknown
parameters, one should always have at least ($n+1$) data points for a
robust and stable fit, and in all our CCM calculations in practice we
try our best to obey this primary edict, in so far as it is possible
to do so with the available computing power. In so far as is possible
we also try to avoid using the least approximate data points (e.g.,
LSUB$m$, SUB$m$-$m$, DSUB$m$ points with $m \leq 2$) since these
low-$m$ data points are rather far from the corresponding large-$m$
limits. In the ensuing discussion we refer to this as our secondary
edict. Nevertheless, we do include such points if it is necessary to
do so to preserve our above primary edict. In these latter cases,
however, we are always careful to do some other careful consistency
checks on the robustness and accuracy of our results.

In the next two sections we now illustrate the use and power of the
new LPSUB$m$ scheme by applying it to two prototypical spin-half
models defined on the 2D square lattice, namely the $XXZ$ model in
section \ref{XXZ} and the $XY$ model in section \ref{XY}.

\section{The spin-$1/2$ antiferromagnetic $XXZ$ model on the square lattice}
\label{XXZ}
As an illustration of the use of the LPSUB$m$ scheme we first consider
its application to the spin-1/2 $XXZ$ model on the infinite square
lattice. The Hamiltonian of the $XXZ$ model, in global spin
coordinates, is written as
\begin{equation}
H_{XXZ} = \sum_{\langle i,j \rangle}[s^{x}_{i}s^{x}_{j} + s^{y}_{i}s^{y}_{j}
+ \Delta s^{z}_{i}s^{z}_{j}], \label{eq:H_XXZ}
\end{equation}
where the sum on $\langle i,j \rangle$ runs over all nearest-neighbour
pairs of sites on the lattice and counts each pair only once. Since
the square lattice is bipartite, we consider $N$ to be even, so that
each sublattice contains $\frac{1}{2}N$ spins, and we consider only the case
where $N \rightarrow \infty$. The N\'{e}el state is the ground state
(GS) in the trivial Ising limit $\Delta \rightarrow \infty$, and a
phase transition occurs at $\Delta = 1$. Indeed, the
classical GS demonstrates perfect N\'{e}el order in the $z$-direction
for $\Delta > 1$, and a similar perfectly ordered $x$-$y$ planar
N\'{e}el phase for $-1 < \Delta < 1$. For $\Delta < -1$ the classical
GS is a ferromagnet.

The case $\Delta = 1$ is equivalent to the isotropic Heisenberg model,
whereas $\Delta = 0$ is equivalent to the isotropic version of the
$XY$ model considered in section~\ref{XY} below. The $z$ component of total
spin, $s^{z}_{T}$, is a good quantum number as it commutes with the
Hamiltonian of equation (\ref{eq:H_XXZ}). Thus one may readily check
that $[s^{z}_{T},H_{XXZ}]=0$. Our interest here is in those values of
$\Delta$ for which the GS is an antiferromagnet.

The CCM treatment of any spin system is initiated by choosing an
appropriate model state $|\Phi\rangle$ (for a particular regime), so
that a linear combinations of products of spin-raising operators can
be applied to this state and all possible spin configurations are
determined. There is never a unique choice of model state
$|\Phi\rangle$. Clearly our choice should be guided by any physical
insight that we can bring to bear on the system or, more specifically,
to that particular phase of it that is under consideration. In the
absence of any other insight into the quantum many-body system it is
common to be guided by the behaviour of the corresponding classical
system (i.e., equivalently, the system when the spin quantum number $s
\rightarrow \infty$). The $XXZ$ model under consideration provides
just such an illustrative example. Thus, for $\Delta > 1$ the {\it
  classical} Hamiltonian of equation (\ref{eq:H_XXZ}) on the 2D square
lattice (and, indeed, on any bipartite lattice) is minimized by a
perfectly antiferromagnetically N\'{e}el-ordered state in the spin
$z$-direction. However, the classical gs energy is minimized by a
N\'{e}el-ordered state with spins pointing along any direction in the
spin $x$-$y$ plane (say, along the spin $x$-direction) for $-1 <
\Delta < 1$. Either of these states could be used as a CCM model state
$|\Phi\rangle$ and both are likely to be of value in different regimes
of $\Delta$ appropriate to the particular quantum phases that mimic
the corresponding classical phases. For present illustrative purposes
we restrict ourselves to the $z$-aligned N\'{e}el state as our choice
for $|\Phi\rangle$, written schematically as $|\Phi\rangle=|\cdots
\downarrow \uparrow \downarrow \uparrow
\cdots\rangle$, in the global spin axes,
where $|\uparrow\rangle\equiv \bigl\vert
\frac{1}{2},+\frac{1}{2}\bigl\rangle$ and $|\downarrow\rangle\equiv
\bigl\vert \frac{1}{2},-\frac{1}{2}\bigl\rangle$ in the usual $|s,
m_{s}\rangle$ notation. Such a state is, clearly, likely to be a good
starting-point for all $\Delta > 1$, down to the expected phase
transition at $\Delta=1$ from a $z$-aligned N\'{e}el phase to an
$x$-$y$ planar N\'{e}el phase.

As indicated in section~\ref{spin_latt} it is now convenient to perform a
rotation of the axes for the up-pointing spins (i.e., those on the
sublattice with spins in the positive $z$-direction) by $180^{\circ}$
about the spin $y$-axis, so that $|\Phi\rangle$ takes the form given by
equation (\ref{local_quan}). Under this rotation, the spin operators
on the original up sub-lattice are transformed as
\begin{equation}
s^{x}\rightarrow-s^{x}, \qquad s^{y} \rightarrow s^{y}, \qquad s^{z} \rightarrow -s^{z}.
\end{equation}
The Hamiltonian of equation (\ref{eq:H_XXZ}) may thus be rewritten in
these local spin coordinate axes as
\begin{equation}
H_{XXZ} = -\frac{1}{2}\sum_{\langle i,j \rangle}[s^{+}_{i}s^{+}_{j} + s^{-}_{i}s^{-}_{j} + 2\Delta s^{z}_{i}s^{z}_{j}]. \label{eq:Ham_XXZ_trans}
\end{equation}

As in any application of the CCM to spin-lattice systems, we now
include in our approximations at any given order only those {\it
  fundamental configurations} that are distinct under the point and
space group symmetries of both the lattice and the Hamiltonian. The
number, $N_{f}$, of such fundamental configurations at any level of
approximation may be further restricted whenever additional
conservation laws come into play. For example, in our present case,
the $XXZ$ Hamiltonian of equation ({\ref{eq:H_XXZ}) commutes with the
  total uniform magnetization, $s^{z}_{T}=\sum^{N}_{k=1}s^{z}_{k}$, in
  the global spin coordinates, where $k$ runs over all lattice
  sites. The GS is known to lie in the $s^{z}_{T}=0$ subspace, and
  hence we exclude configurations with an odd number of spins or with
  unequal numbers of spins on the two equivalent sublattices of the
  bipartite square lattice. We show in figure~\ref{LPSUB1to3_xxz}
\begin{figure}[!t]
\begin{center}                  
\includegraphics[width=10cm]{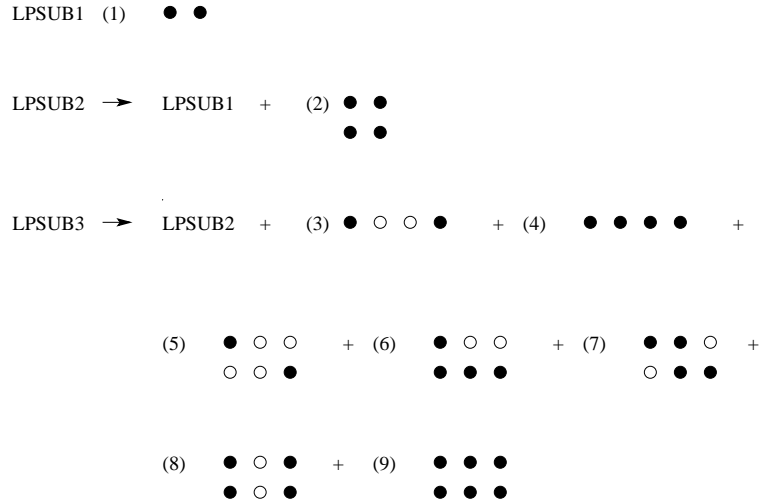} 
\vskip0.5cm
\caption{The fundamental configurations for the LPSUB$m$ scheme with
  $m=\{1,2,3\}$ for the spin-1/2 $XXZ$ model on a square lattice in
  two dimensions. The filled circles mark the relative positions of
  the sites of the square lattice on which the spins are flipped with
  respect to the model state. The unfilled circles represent unflipped
  sites.}
\label{LPSUB1to3_xxz}                                                       
\end{center}
\end{figure}   
the fundamental configurations that are accordingly allowed for the
LPSUB$m$ approximations for this spin-1/2 $XXZ$ model on the 2D square
lattice, with $1 \leq m \leq 3$. We see, for example, that $N_{f}=9$
at the LPSUB3 level of approximation. 

The LPSUB$m$ approximations can readily be implemented for the present
spin-1/2 $XXZ$ model on the 2D square lattice for all values $m \leq
6$ with reasonably modest computing power. By comparison, the LSUB$m$
and DSUB$m$ schemes can both be implemented with comparable computing
resources for all values $m \leq 9$. Numerical results for the gs
energy per spin and the sublattice magnetization are shown in
table~\ref{LPSUBm_table_xxz} at the isotropic point $\Delta=1$ at
various levels of approximation, and corresponding results for the
same quantities are displayed graphically in figures
\ref{LPSUBm_XXZ_E} and \ref{LPSUBm_XXZ_M} as functions of the
anisotropy parameter $\Delta$.
\begin{table}[!tbh]    	
\begin{center}   
  \caption{\label{LPSUBm_table_xxz} The ground-state energy per spin
    ($E/N$) and sublattice magnetization ($M$) for the spin-$1/2$
    $XXZ$ model on the 2D square lattice, obtained using the CCM
    LPSUB$m$ approximation scheme with $1 \leq m \leq 6$ at
    $\Delta=1$. $N_{f}$ is the number of fundamental configurations at
    a given LPSUB$m$, LSUB$m$ or DSUB$m$ level of
    approximation. $\Delta_{i} \equiv$ LPSUB$m$ sublattice
    magnetization point of inflexion. The LPSUB$m$ results for odd
    values of $m$, even values of $m$ and the whole series of $m$ are
    extrapolated separately. These results are compared to
    calculations using third-order spin-wave theory
    (SWT)~\cite{Ha:1992}, series expansions (SE)~\cite{Zh:1991}, exact
    diagonalization (ED)~\cite{Rit:2004}, quantum Monte Carlo
    (QMC)~\cite{Sa:1997}, LSUB$\infty$ extrapolations of the CCM
    LSUB$m$ approximations~\cite{Bi:2000,Ri:2008,Fa:2008} and the
    DSUB$\infty$ extrapolations of the CCM DSUB$m$
    approximations~\cite{Bi:2009_DSUBm}.}
\begin{tabular}{ccccccc}
\br
{Method} & {$N_{f}$} & {$E/N$} & {$M$} & {$\Delta_{i}$} & {$\Delta_{c}$} & Max. no. \\  \cline{3-4}   
&  & \multicolumn{2}{c}{$\Delta = 1$} & &  & of spins  \\  
\mr
LPSUB1 & 1 & -0.64833 & 0.421 &  &  &  2 \\ 
LPSUB2 & 2 & -0.65311 & 0.410 & 0.258 &  & 4 \\  
LPSUB3 & 9 & -0.66442 & 0.379 &  & 0.579 & 6 \\ 
LPSUB4 & 35 & -0.66565 & 0.372 & 0.586 &  & 8 \\ 
LPSUB5 & 265 & -0.66761 & 0.358 &  & 0.766 & 12  \\ 
LPSUB6 & 2852 & -0.66807 & 0.354 &  0.735 &  & 16  \\ 
\mr
LSUB8 & 1287 & -0.66817 & 0.352 &  & 0.844 & 8  \\ 
LSUB10 & 29605 & -0.66870 & 0.345 &  &  & 10 \\ 
\br
& \multicolumn{5}{c}{Extrapolation} & Based on \\ 
\mr
LPSUB$\infty$ & & -0.66953 & 0.320 &  & & $m=\{1,3,5\}$ \\ 
LPSUB$\infty$ & & -0.67004 &  0.308 & 1.093 &  & $m=\{2,4,6\}$ \\  
LPSUB$\infty$ & & -0.66867 & 0.328 &  &  & $2 \leq m \leq 6$ \\ 
LPSUB$\infty$ & & -0.67107 & 0.288  &  &  & $3 \leq m \leq 6$ \\  
\mr
DSUB$\infty$ & & -0.67082 & 0.308 & 1.009 &  & $m=\{6,8,10\}$ \\   
\mr
LSUB$\infty$ & & -0.67029 & 0.304 & & & $n=\{3,5,7,9\}$ \\ 
LSUB$\infty$ & & -0.66966 & 0.310 & & & $n=\{4,6,8,10\}$ \\ 
LSUB$\infty$ & & -0.66962 & 0.308 & & & $n=\{6,8,10\}$ \\  
\br
SWT &  & -0.66999 & 0.3069 & & \\    
SE & & -0.66930 & 0.307 & & \\
ED& & -0.67000 & 0.317 &  \\ 
QMC & & -0.669437(5) & 0.3070(3) & & \\      
\br
\end{tabular}   
\end{center}
\end{table} 
\begin{figure}[!b]  
\begin{center}
\includegraphics[width=8cm]{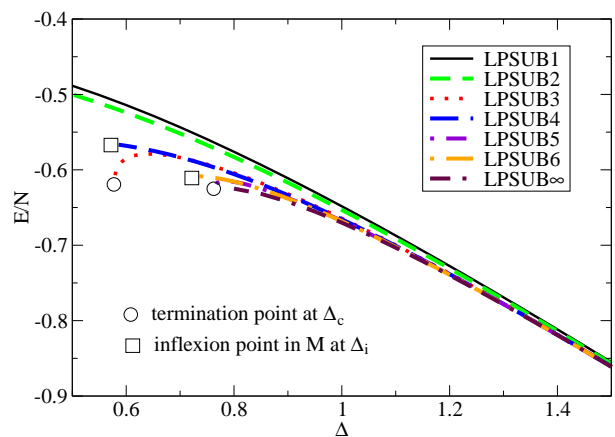} 
\caption{(Color online) CCM results for the ground-state energy per
  spin, $E/N$, as a function of the anisotropy parameter $\Delta$, of
  the spin-$1/2$ $XXZ$ model on the 2D square lattice, using various
  LPSUB{\it m} approximations based on the {\it z}-aligned N\'{e}el
  model state. The LPSUB$m$ results with $m=\{2,4,6\}$ are
  extrapolated using the quadratic fit of equation
  (\ref{Extrapo_E}) and shown as the curve
  LPSUB$\infty$. $\Delta_{i}\equiv$ magnetization point of inflexion,
  described in the text.}
\vskip0.1cm
\label{LPSUBm_XXZ_E}
\end{center}
\end{figure}
\begin{figure}[!t] 
\begin{center}
\vskip0.5cm
\includegraphics[width=8cm]{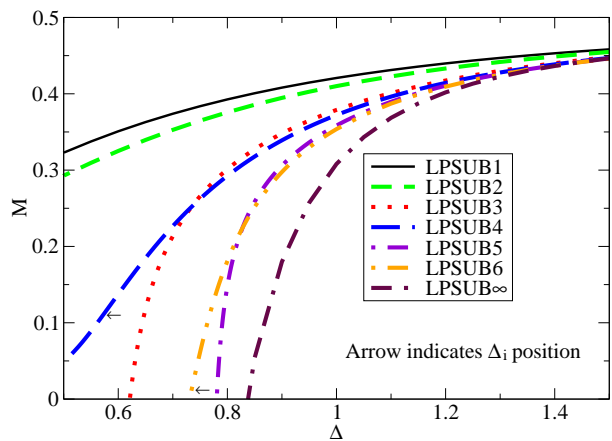}  
\caption{(Color online) CCM results for the ground-state sublattice
  magnetization, $M$, as a function of the anisotropy parameter
  $\Delta$, of the spin-$1/2$ $XXZ$ model on the 2D square lattice, using
  various LPSUB{\it m} approximations based on the {\it z}-aligned N\'{e}el
  model state. The LPSUB$m$ results with $m=\{2,4,6\}$ are
  extrapolated using the quadratic fit of equation
  (\ref{Extrapo_M}) and shown as the curve
  LPSUB$\infty$. $\Delta_{i}\equiv$ point of inflexion in the curve,
  shown by arrows in the figure.}
\label{LPSUBm_XXZ_M}
\end{center}
\end{figure}

We also show in table~\ref{LPSUBm_table_xxz} for the isotropic
Heisenberg Hamiltonian ($\Delta=1$) the results for the gs energy and
sublattice magnetization using the (quadratic) extrapolation schemes
of equations (\ref{Extrapo_E}) and (\ref{Extrapo_M}) respectively of
the LPSUB$m$ data, employing various subsets of results. Comparison is
also made with corresponding LSUB$m$~\cite{Fa:2008,Ri:2008} and
DSUB$m$~\cite{Bi:2009_DSUBm} extrapolation schemes for the same
model. The results are generally observed to agree very well 
with each other. Excellent agreement of all the CCM extrapolations is
also obtained with the results from the best of the alternative
methods for this model, including third-order spin-wave theory
(SWT)~\cite{Ha:1992}, linked-cluster series expansion (SE)
techniques~\cite{Zh:1991}, the extrapolations to infinite lattice size
($N \rightarrow \infty$) from the exact diagonalization (ED) of small
lattices~\cite{Rit:2004}, and quantum Monte Carlo (QMC) calculations
for larger lattices~\cite{Sa:1997}.

As discussed in section \ref{Extrapo} we always prefer to have at
least 4 LPSUB$m$ calculations with different values of the truncation
index $m$, to fit to the three unknown parameters of the quadratic
fitting expressions for $E/N$ and $M$ in equations (\ref{Extrapo_E})
and (\ref{Extrapo_M}). This primary edict is not violated if we
extrapolate the LPSUB$m$ data using both odd and even values of $m$ in
the range $2 \leq m \leq 6$ or $3 \leq m \leq 6$. We note, however,
that if we extrapolate using only the three even values $m=\{2,4,6\}$
or using the three odd values $m=\{1,3,5\}$ then we violate both the
primary and secondary edicts discussed above. Nevertheless, the
extrapolated results using the even set $m=\{2,4,6\}$ are seen to be
in good agreement with those from the alternative methods shown in
table \ref{LPSUBm_table_xxz}.

It has been observed and well documented in the past (and see, e.g.,
Ref.~\cite{Fa:2008}) that the CCM LSUB$m$ results for this model (and
many others) for both the gs energy $E$ and the sublattice
magnetization $M$ show a distinct period-2 ``staggering'' effect with
index $m$, according to whether $m$ is even or odd. As a consequence
the LSUB$m$ data for both $E$ and $M$ converge differently for the
even-$m$ and the odd-$m$ sequences. This is very similar to what is also
observed very frequently in perturbation theory in corresponding
even and odd orders~\cite{Mo:1953}. As a rule, therefore, the LSUB$m$
data are generally extrapolated separately for even $m$ and for odd
values of $m$, since the staggering makes extrapolations using both
odd and even values together rather difficult. We show in
figure~\ref{LPSUBm_staggered_all}
\begin{figure}[!t]
\begin{center}
\mbox{
  \subfigure[Ground-state energy per spin]{\label{LPSUBm_E_staggered_XXZ}\includegraphics[scale=0.3,angle=270]{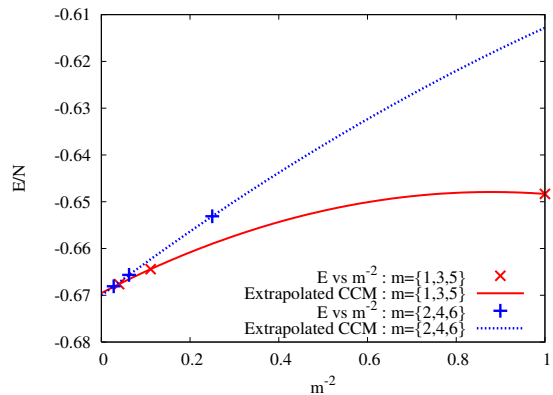}} 
}
\mbox{
  \subfigure[Ground-state sublattice magnetization]{\label{LPSUBm_M_staggered_XXZ}\includegraphics[scale=0.3,angle=270]{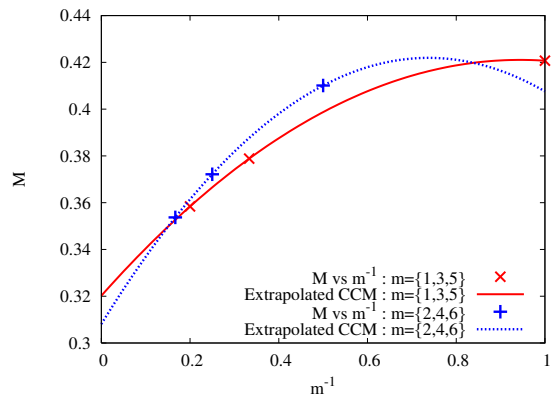}} 
}
\caption{(Color online) Illustration of the odd-even staggered nature
  with respect to the truncation parameter $m$ of the LPSUB$m$ scheme
  results for the ground-state energy per spin, $E/N$, and sublattice
  magnetization, $M$, for the spin-$1/2$ antiferromagnetic $XXZ$ model
  on the 2D square lattice, for the isotropic limiting case
  $\Delta=1$. The LPSUB$m$ data are plotted against $1/m^{2}$ for
  $E/N$ and against $1/m$ for $M$. The results clearly justify the
  heuristic extrapolation schemes of equations (\ref{Extrapo_E}) and
  (\ref{Extrapo_M}).}
\label{LPSUBm_staggered_all}
\end{center}
\end{figure}         
our LPSUB$m$ results for the gs energy per spin and the sublattice
magnetization plotted against $1/m^{2}$ and $1/m$, respectively, for
the case $\Delta=1$. The higher odd and even $m$ values taken together
clearly cluster well in both cases on straight lines, thereby
justifying {\it a posteriori} our heuristic extrapolation fits of
equations (\ref{Extrapo_E}) and (\ref{Extrapo_M}). Just as in the
LSUB$m$ case a small but definite ``odd-even staggering'' effect is
observed in the LPSUB$m$ data for both the energy and the sublattice
magnetization, although it is less pronounced than for the
corresponding DSUB$m$~\cite{Bi:2009_DSUBm} and LSUB$m$
data~\cite{Fa:2008} for this model.

Before discussing our LPSUB$m$ results further for this model we note
that the comparable LSUB$m$ solutions actually terminate at a critical
value $\Delta_{c}=\Delta_{c}(m)$, which itself depends on the
truncation index $m$~\cite{Fa:2004}. Such LSUB$m$ termination points
are very common for many spin-lattice systems. They have been very
well documented and their origin is clearly understood (and see, e.g.,
Ref.~\cite{Fa:2004}). Thus, in all such cases a termination point
always arises due to the solution of the CCM equations becoming
complex at this point, beyond which there exist two branches of
entirely unphysical complex conjugate solutions~\cite{Fa:2004}. In the
region where the solution reflecting the true physical solution is
real there actually also exists another (unstable) real
solution. However, only the (shown) upper branch of these two
solutions reflects the true (stable) physical GS, whereas the lower
branch does not. The physical branch is usually easily identified in
practice as the one which becomes exact in some known (e.g.,
perturbative) limit. This physical branch then meets the corresponding
unphysical branch at the corresponding termination point, beyond which
no real solutions exist. The LSUB$m$ termination points are themselves
also reflections of the quantum phase transitions in the real system,
and may hence be used to estimate the position of the phase
boundary~\cite{Fa:2004}.

We note that when the LPSUB$m$ approximations are applied to the $XXZ$
model, only the odd LPSUB$m$ levels with $m \geq 3$ terminate in the
same way as do the corresponding LSUB$n$ approximations, as shown in
figures~\ref{LPSUBm_XXZ_E} and \ref{LPSUBm_XXZ_M}. The LPSUB$m$
solutions with even values $m=\{2,4,6\}$ do not terminate. We have no
convincing explanation for this difference in behaviour for two
apparently similar schemes applied to the same model. Nevertheless, it
is still possible to use our LPSUB$m$ data to extract an estimate for
the physical phase transition point at which the $z$-aligned N\'{e}el
phase terminates. As has been justified and utilized
elsewhere~\cite{Fa:1997}, a point of inflexion at $\Delta=\Delta_{i}$
in the sublattice magnetization $M$ as a function of $\Delta$ also
indicates the onset of an instability in the system. Such inflexion
points $\Delta_{i}=\Delta_{i}(m)$ occur for the even values of the
LPSUB$m$ approximations, as indicated in table~\ref{LPSUBm_table_xxz}
and figure~\ref{LPSUBm_XXZ_M}. The LPSUB$m$ approximations are thus
expected to be unphysical for $\Delta < \Delta_{i}(m)$, and we hence
show the corresponding results for the gs energy per spin in
figure~\ref{LPSUBm_XXZ_E} only for values $\Delta_{i} >
\Delta_{i}(m)$. Heuristically, we find that the magnetization
inflexion points $\Delta_{i}(m)$ scale linearly with $(1/m)$ in the
large $m$ limit, and the extrapolated results shown in
table~\ref{LPSUBm_table_xxz} have been performed with,
$\Delta_{i}=c_{o}+c_{1}(1/m)+c_{2}(1/m)^{2}$, commensurate with the
corresponding fits in $1/m^{2}$ and $1/m$ for the gs energy per spin
and sublattice magnetization of equations (\ref{Extrapo_E}) and
(\ref{Extrapo_M}), respectively. The extrapolated values from both the
LPSUB$m$ and DSUB$m$ schemes are in excellent agreement with the
expected phase transition point at $\Delta_{c} \equiv 1$ between two
quasiclassical N\'{e}el-ordered phases aligned along the spin $z$-axis
(for $\Delta > 1$) and in some arbitrary direction in the spin
$x$-$y$-plane (for $|\Delta| < 1$).

Although we do not do so here, the $x$-$y$ planar N\'{e}el phase could
itself also easily be investigated by another CCM LPSUB$m$ series of
calculations based on a model state $|\Phi\rangle$ with perfect
N\'{e}el ordering in, say, the $x$-direction. Nevertheless,
from our results so far we observe that the LPSUB$m$ scheme has
at, least partially, fulfilled the expectations placed upon it for the
present model. Accordingly, we now apply it to the second test model
of the spin-1/2 $XY$ model on the 2D square lattice.

\section{The spin-1/2 $XY$ model on the square lattice}
\label{XY}
The Hamiltonian of the $XY$ model~\cite{Fa:1997} in global spin
coordinates, is written as
\begin{equation}
H_{XY} = \sum_{\langle i,j \rangle}[(1+\Delta)s^{x}_{i}s^{x}_{j} + (1-\Delta)s^{y}_{i}s^{y}_{j}];\hspace{0.2in} -1 \leq \Delta \leq 1,  \label{eq:H_XY}
\end{equation}
where the sum on $\langle i,j \rangle$ again runs over all
nearest-neighbour pairs of lattice sites and counts each pair only
once. We again consider the case of spin-1/2 particles on each site of
an infinite 2D square lattice.

For the classical model described by equation (\ref{eq:H_XY}), it is
clear that the GS is a N\'{e}el state in the $x$-direction for $0 <
\Delta \leq 1$ and a N\'{e}el state in the $y$-direction for $-1 \leq
\Delta < 0$. Hence, since we only consider the case $0 \leq \Delta
\leq 1$, we choose as our CCM model state $|\Phi\rangle$ for the
quantum $XY$ model a N\'{e}el state aligned along the $x$-direction,
written schematically as, $|\Phi\rangle=|\cdots \leftarrow \, \rightarrow \, \leftarrow \, \rightarrow
\cdots\rangle,$\ in the global spin axes.
Clearly the case $-1 \leq \Delta < 0$ is readily obtained from the case $0
< \Delta \leq 1$ by interchange of the $x$- and $y$-axes.

As before in section \ref{XXZ} we now perform our usual rotation of the spin axes on each
lattice site so that $|\Phi\rangle$ takes the form given by equation
(\ref{local_quan}}) in the rotated local spin coordinate frames. Thus,
for the spins on the sublattice where they point in the negative
$x$-direction in the global spin axes (i.e., the left-pointing spins)
we perform a rotation of the spin axes by $+90^{\circ}$ about the spin
$y$-axis. Similarly, for the spins on the other sublattice where they
point in the positive $x$-direction in the global spin axes (i.e., the
right-pointing spins) we perform a rotation of the spin axes by
$-90^{\circ}$ about the spin $y$-axis. Under these rotations the spin
operators are transformed as
\numparts
\begin{eqnarray}    
s^{x} \rightarrow  s^{z}\;, \qquad s^{y}  \rightarrow s^{y},  \qquad s^{z} \rightarrow -s^{x}, \qquad \mbox{left-pointing spins}; \\
s^{x} \rightarrow -s^{z}\;, \qquad s^{y}  \rightarrow s^{y}, \qquad  s^{z} \rightarrow s^{x}, \qquad  \mbox{right-pointing spins}.
\end{eqnarray} 
\endnumparts
The Hamiltonian of equation (\ref{eq:H_XY}) may thus be rewritten in the local spin coordinate axes defined above as
\begin{equation}
\fl H_{XY} = \sum_{\langle i,j \rangle}\left[-(1+\Delta)s^{z}_{i}s^{z}_{j}-\frac{1}{4}(1-\Delta)(s^{+}_{i}s^{+}_{j}+s^{-}_{i}s^{-}_{j}) 
+\frac{1}{4}(1-\Delta)(s^{+}_{i}s^{-}_{j}+s^{-}_{i}s^{+})\right]. \label{Hamilton_XY_trans}
\end{equation}



\begin{figure}[!t]    
\begin{center}
\includegraphics[width=10cm]{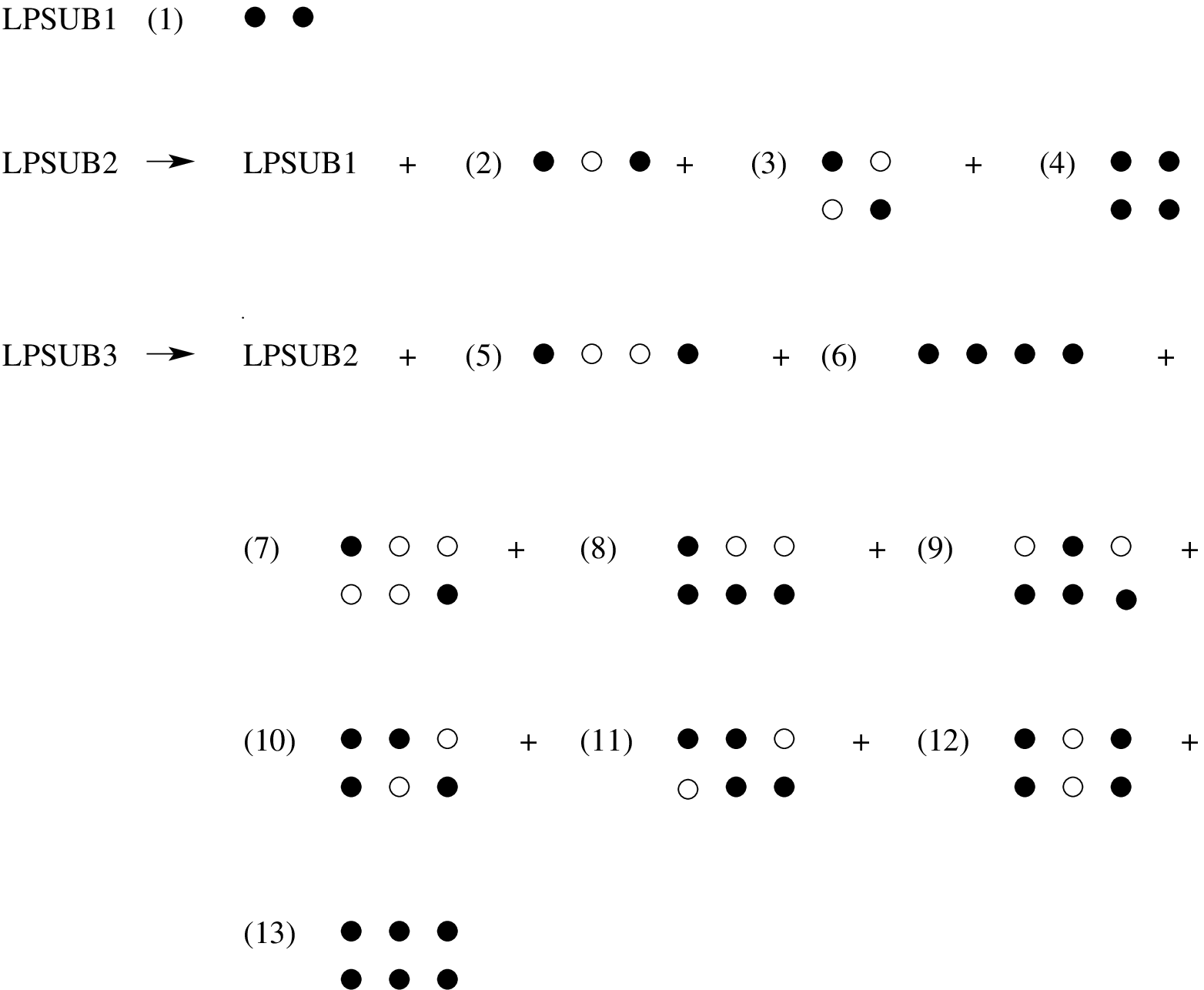} 
\vskip0.5cm
\caption{The fundamental configurations for the LPSUB$m$ scheme with
  $m=\{1,2,3\}$ for the spin-1/2 $XY$ model on a square lattice in two dimensions. The filled circles mark the
  relative positions of the sites of the square lattice on which the
  spins are flipped with respect to the model state.  The unfilled
  circles represent unflipped sites.}
\label{LPSUB1to3_xy}
\end{center}
\end{figure}

Exactly as in the previous application, we now have to evaluate the
fundamental configurations that are retained in the CCM correlation
operators $S$ and $\tilde{S}$ at each LPSUB$m$ level of
approximation. Although the point and space group symmetries of the
square lattice (common to both the $XXZ$ and $XY$ models considered
here) and the two Hamiltonians of equations (\ref{eq:Ham_XXZ_trans})
and (\ref{Hamilton_XY_trans}) are identical, the numbers $N_{f}$ of
fundamental configurations for a given LPSUB$m$ level are now larger
(except for the case $m=1$) for the $XY$ model than for the $XXZ$
model, since the uniform magnetization is no longer a good quantum
number for the $XY$ model, $[H_{XY},s^{z}_{T}]\neq 0$. Nevertheless,
we note from the form of equation (\ref{Hamilton_XY_trans}), in which
the spin-raising and spin-lowering operators appear only in
combinations that either raise or lower the number of spin flips by
two (viz., the $s^{+}_{i}s^{+}_{j}$ and $s^{-}_{i}s^{-}_{j}$
combinations, respectively) or leave them unchanged (viz., the
$s^{+}_{i}s^{-}_{j}$ and $s^{-}_{i}s^{+}_{j}$ combinations), it is
only necessary for the $s^{z}_{T}=0$ GS to consider fundamental
configurations that contain an even number of spins. Thus, the main
difference for the $XY$ model over the $XXZ$ model is that we must now
also consider fundamental configurations in which we drop the
restriction for the former case of having an equal number of spins on
the two equivalent sublattices of the bipartite square lattice that
was appropriate for the latter case. We show in
figure~\ref{LPSUB1to3_xy} the fundamental configurations that are
allowed for the spin-1/2 $XY$ model on the square lattice for the
LPSUB$m$ approximation with $1 \leq m \leq 3$, and we invite the
reader to compare with the corresponding fundamental configurations
for the spin-1/2 $XXZ$ model on the same square lattice shown in
figure 1.
The corresponding numbers $N_{f}$ of fundamental configurations for
the $XY$ model are also shown in table~\ref{table_LPSUBm_XY} for the
LPSUB$m$ approximations with $m \leq 6$.

\begin{figure}[!t]
\begin{center}
\vskip0.2cm
\includegraphics[width=8cm]{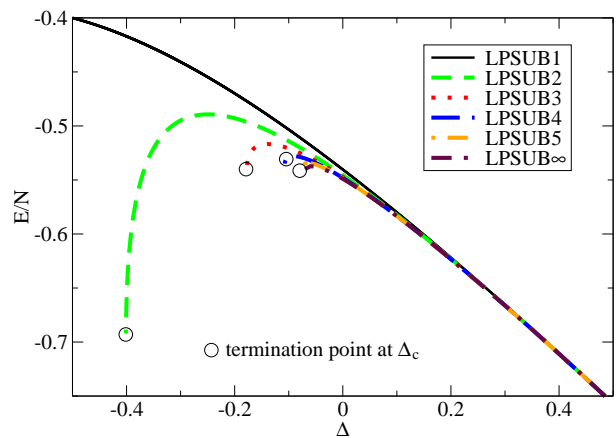}
\caption{(Color online) CCM results for the ground-state energy per
  spin, $E/N$, as a function of the anisotropy parameter $\Delta$, of
  the spin-$1/2$ $XY$ model on the 2D square lattice obtained using
  the LPSUB{\it m} approximation based on the N\'{e}el state aligned
  along any axis in the $x$-$y$ plane.  The LPSUB$m$ results with
  $m=\{1,3,5\}$ are extrapolated using equation (\ref{Extrapo_E}) to
  give the curve labelled DSUB$\infty$.}
\vskip0.2cm
\label{LPSUBm_XY_E}
\end{center}
\end{figure}

\begin{figure}[!t]
\begin{center}
\includegraphics[width=8cm]{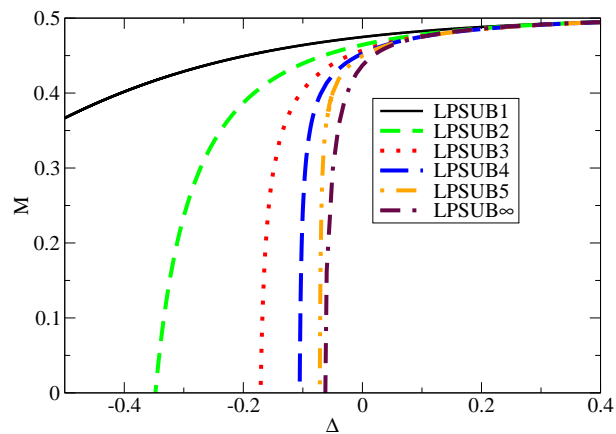}
\caption{(Color online) CCM results for the ground-state sublattice
  magnetization, $M$, as a function of the anisotropy parameter
  $\Delta$, of the spin-$1/2$ $XY$ model on the 2D square lattice
  obtained using various LPSUB{\it m} approximations based on the
  N\'{e}el state aligned along any axis in the $x$-$y$ plane.  The
  LPSUB$m$ results with $m=\{1,3,5\}$ are extrapolated using equation
  (\ref{Extrapo_M}) to give the curve labelled DSUB$\infty$.}
\label{LPSUBm_XY_M}
\end{center}
\end{figure}

We present results for the spin-1/2 $XY$ model on the square lattice
in the CCM LPSUB$m$ approximations for all values $m \leq 6$, all of
which can be easily computed with very modest computing
power. Comparable computing power enables the corresponding LSUB$m$
scheme to be implemented for all $m \leq 8$. Numerical results for the
gs energy per spin and sublattice magnetization are shown in
table~\ref{table_LPSUBm_XY} at the isotropic point at $\Delta=0$ at
various levels of approximation, and corresponding results for the
same gs quantities are shown graphically in figures~\ref{LPSUBm_XY_E}
and~\ref{LPSUBm_XY_M} as functions of the anisotropy parameter
$\Delta$.
\begin{table}[!tbh]    
\begin{center}       
 \caption{\label{table_LPSUBm_XY}The ground-state energy per spin ($E/N$) and sublattice magnetization ($M$) for the spin-$1/2$ $XY$ model on the 2D square lattice, obtained using the CCM LPSUB$m$ approximation scheme with $1 \leq m \leq 6$ at $\Delta=0$. $N _{f}$ is the number of fundamental configurations at a given level of LPSUB$m$, LSUB$m$ or DSUB$m$ approximation. $\Delta_{c} \equiv$ LPSUB$m$ termination point. The LPSUB$m$ results for odd values of $m$, even values of $m$ and the whole series of $m$ are extrapolated separately. These results are compared to calculations using series expansions (SE)~\cite{Ha:1991}, the quantum Monte Carlo (QMC) method~\cite{Sa:1999}, LSUB$\infty$ extrapolations of the CCM LSUB$m$ approximations~\cite{Fa:1997} and the DSUB$\infty$ extrapolations of the CCM DSUB$m$ approximations~\cite{Bi:2009_DSUBm}.} 
\begin{tabular}{cccccc}
\br
Method & $N_{f}$ & $E/N$ & $M$ & $\Delta_{c}$ & Max. no. \\ \cline{3-4}    
&  & \multicolumn{2}{c}{$\Delta = 0$}  & & of spins  \\  
\mr  
LPSUB1 & 1  & -0.54031 & 0.475 &  $^{a}$ &  2 \\ 
LPSUB2 & 4 & -0.54548 & 0.464 & -0.401 & 4 \\  
LPSUB3 & 13 & -0.54747 & 0.457 & -0.178 & 6\\ 
LPSUB4 & 72 & -0.54812 & 0.453 & -0.107 & 8 \\ 
LPSUB5 & 557 & -0.54842 &  0.450 & -0.072 &  12 \\ 
LPSUB6 & 7410 & -0.54857 & 0.448  &  $^{b}$ & 16  \\  
\mr
LSUB6 &  131 & -0.54833 & 0.451 & -0.073 & 6 \\    
LSUB8 &  2793 & -0.54862 & 0.447 & -0.04 & 8 \\ 
\br
\multicolumn{5}{c}{Extrapolation} & Based on  \\  
\mr
LPSUB$\infty$ & & -0.54894 & 0.437 & -0.017 & $2 \leq m \leq 5$ \\ 
LPSUB$\infty$ & & -0.54897 & 0.435  & -0.006   & $3 \leq m \leq 5$  \\ 
LPSUB$\infty$ & & -0.54893 & 0.436  &  $^{b}$  & $2 \leq m \leq 6$  \\ 
LPSUB$\infty$ & & -0.54894 & 0.435  & $^{b}$ & $3 \leq m \leq 6$  \\ 
LPSUB$\infty$ & & -0.54888 & 0.436  & $^{b}$  & $4 \leq m \leq 6$  \\      
LPSUB$\infty$ & & -0.54899 & 0.437 & $^{a}$ & $m=\{1,3,5\}$ \\ 
LPSUB$\infty$ & & -0.54893 & 0.436  & $^{b}$ & $m=\{2,4,6\}$ \\  
\mr
LSUB$\infty$ & & -0.54892 & 0.435 & 0.00 & $n=\{4,6,8\}$ \\  
\hline\noalign{\smallskip}   
DSUB$\infty$ & & -0.54950 & 0.436 &  & $m=\{3,5,7,9\}$ \\ 
DSUB$\infty$ & & -0.54923 & 0.437 & 0.011 & $m=\{5,7,9\}$  \\ 
\br
SE &  & -0.5488 & 0.436 & 0.0 &  \\    
QMC &   & -0.54882(2) & 0.437(2) & & \\   
\br       
\end{tabular}    
\end{center}
\vskip0.1cm \underline{NOTES}: \protect \\ $^{a}$ The LPSUB1
approximation does not terminate.  \\ $^{b}$ The spin-flip
configurations for the LPSUB6 approximation are sufficiently
complicated and large in number that calculations have only been
calculated at present for $\Delta=0$ for the LPSUB6 case.
\end{table}         

We also show in table~\ref{table_LPSUBm_XY} for the isotropic $XY$
Hamiltonian ($\Delta=0$) the results for the gs energy and sublattice
magnetization using the (quadratic) extrapolation schemes of equations
(\ref{Extrapo_E}) and (\ref{Extrapo_M}) respectively of the LPSUB$m$
data, employing various subsets of our results, just as for the $XXZ$ model
considered previously. We also compare in table~\ref{table_LPSUBm_XY}
the present results with the corresponding CCM LSUB$m$ \cite{Fa:1997}
and DSUB$m$ \cite{Bi:2009_DSUBm} results for the same model. All of
the CCM results are clearly in excellent agreement both with one
another and with the results of the best of the alternative methods
available for this model, including the linked-cluster series
expansion (SE) technique~\cite{Ha:1991} and a quantum Monte Carlo
(QMC) method~\cite{Sa:1999}.

\begin{figure}[!t]
\begin{center}            
\mbox{
	\subfigure[Ground-state energy per spin]{\label{LPSUBm_E_staggered_XY}\includegraphics[scale=0.30,angle=270]{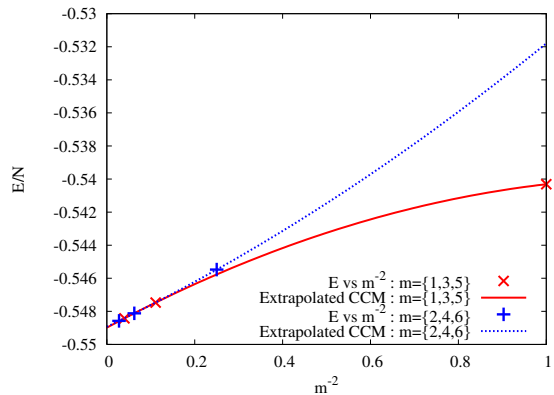}} 
}
\mbox{   
	\subfigure[Ground-state sublattice magnetization]{\label{LPSUBm_M_staggered_XY}\includegraphics[scale=0.30,angle=270]{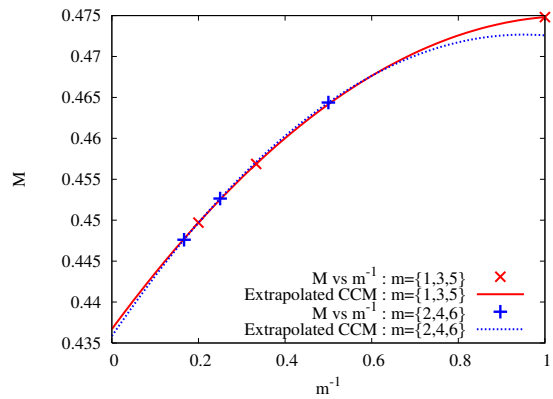}} 
}
\end{center}
\caption{(Color online) Illustration of the odd-even staggered nature
  with respect to the truncation parameter $m$ of the LPSUB$m$ scheme
  results for the ground-state energy per spin, $E/N$, and sublattice
  magnetization, $M$, for the spin-$1/2$ $XY$ model on the 2D square
  lattice, for the isotropic limiting case $\Delta=0$. The LPSUB$m$
  data are plotted against $1/m^{2}$ for $E/N$ and against $1/m$ for
  $M$. The results clearly justify the heuristic extrapolation schemes
  of equations (\ref{Extrapo_E}) and (\ref{Extrapo_M}).}
\label{LPSUBm_staggered_XY_all}
\end{figure}

We show in figure~\ref{LPSUBm_staggered_XY_all} our LPSUB$m$
results for the present $XY$ model for the gs energy per spin and the
sublattice magnetization, plotted respectively against $1/m^{2}$
and $1/m$, for the case $\Delta=0$.
As previously for the $XXZ$ model, the higher $m$ values cluster well
on straight lines in both cases, thereby justifying once more our
heuristic choice of extrapolation fits indicated in equations
(\ref{Extrapo_E}) and
(\ref{Extrapo_M}). Figures~\ref{LPSUBm_E_staggered_XY} and
\ref{LPSUBm_M_staggered_XY} once more show an ``odd-even'' staggering
effect in the termination index $m$ for the LPSUB$m$ data and we have
again shown separate extrapolations of our LPSUB$m$ results in
table~\ref{table_LPSUBm_XY} for the even-$m$ data and the odd-$m$
data, as well as results using all (higher) values of $m$. It is
interesting to note, however, that the staggering effect for this $XY$
model is far less pronounced than for the similar $XXZ$ model in
section \ref{XXZ}. We have no compelling argument to explain this
difference.

It is interesting to note that for the present $XY$ model the CCM
LPSUB$m$ solutions (with our choice of model state as a N\'{e}el state
in the $x$-direction) now do physically terminate for all values of
the truncation index $m \geq 1$ at a critical value
$\Delta_{c}=\Delta_{c}(m)$, exactly as commonly occurs (as for the
present model) for the LSUB$m$ calculations, as we explained above in
section~\ref{XXZ}. Why such LPSUB$m$ terminations occur for all values
$m > 1$ for the $XY$ model but not for odd values of $m$ for the
previous $XXZ$ model is not obvious to us. The corresponding
termination points, $\Delta_{c}=\Delta_{c}(m)$, at various LPSUB$m$,
LSUB$m$ and DSUB$m$ levels of approximation are shown in
table~\ref{table_LPSUBm_XY}. It has been shown
previously~\cite{Bi:1994} that $\Delta_{c}(m)$ scales well with
$(1/m)^{2}$ for the LSUB$m$ data, and the LSUB$\infty$
result~\cite{Fa:1997} shown in table~\ref{table_LPSUBm_XY} was
obtained by the scaling law,
$\Delta_{c}(m)=d_{0}+d_{1}(1/m)^{2}+d_{2}(1/m)^{4}$. We find
heuristically that the best large-$m$ asymptotic behaviour of the
LPSUB$m$ data for $\Delta_{c}(m)$ is also against $(1/m)^{2}$ as the
scaling parameter. Accordingly, the LPSUB$\infty$ values for
$\Delta_{c}$ in table~\ref{table_LPSUBm_XY} are obtained with the same
(quadratic) fit,
$\Delta_{c}(m)=d_{0}+d_{1}(1/m)^{2}+d_{2}(1/m)^{4}$. We see that both
the LSUB$\infty$ and LPSUB$\infty$ results for $\Delta_{c} \equiv
\Delta_{c}(\infty)$ agree very well with the value $\Delta_{c}=0$ that
is known to be the correct value for the phase transition in the
one-dimensional spin-1/2 $XY$ chain from the known exact
solution~\cite{Li:1961}, and which is believed on symmetry grounds
also to be the phase transition point for higher dimensions, including
the present 2D square lattice.

\section{Conclusions}
\label{discussion}
From the two nontrivial benchmark spin-lattice problems that we have
investigated here, it is clear that the new LPSUB$m$ approximation
scheme works well for calculating their gs properties and phase
boundaries. We have utilized here only the simplest 
extrapolation schemes in the pertinent scaling variables, and have
shown that these may be chosen, for example, as $1/m^{2}$ for the
gs energy and $1/m$ for the order parameter. For further use of
the scheme for more complex lattice models (e.g., those exhibiting
geometric or dynamic frustration) it will be necessary to re-visit the
validity of these expansions, but a great deal of previous experience
in such cases for the LSUB$m$ scheme should provide good guidance.

On the basis of the test results presented here, the LPSUB$m$ scheme
clearly does not fulfill the first of our two main criteria for
introducing it, since the number of fundamental configurations,
$N_{f}$, actually increases even more rapidly with truncation index
$m$ than for the corresponding LSUB$m$ series of
approximations. Nevertheless, our second criterion of capturing the
physically most important configurations at relatively low levels of
approximation does seem to be fulfilled, according to our experience
with the convergence of the LPSUB$m$ sequences for observable
quantities. At the very least we now have three schemes (LSUB$m$,
DSUB$m$ and LPSUB$m$) available to us for future investigations of
more complicated spin-lattice models, each of which has its own
merits, and which thus allows us more freedom in future applications
of the CCM to quantum magnetism.

It is particularly worth noting too that our preliminary calculations
have shown that the different schemes show markedly varying patterns
of odd-even staggering, both for a given scheme applied to different
models and for different schemes applied to the same model. It is
difficult to predict in advance how strong or weak the effect will be
for a given scheme applied to a specific model. Nevertheless, when the
effect is weak one can confidently extrapolate the results using both
odd and even values of the truncation parameter $m$ simultaneously,
thereby effectively doubling the number of data points for the fit. In
such cases our first criterion for an improved scheme has effectively
been realized over one where the staggering effect is much more
pronounced, even though the number of fundamental configurations,
$N_{f}$, may indeed increase {\it more} rapidly with truncation index
$m$ for the former (``improved'') scheme than for the latter.

\section*{Acknowledgement}
We are grateful to Dr. J. Schulenburg of Universit\"{a}t Magdeburg for
his assistance in the incorporation into the CCM computer code of the
new LPSUB$m$ approximation scheme.

\end{document}